\documentclass{elsart}
\usepackage{graphicx,amssymb}
\begin{document}

\begin{frontmatter}



\title{Binding energy of an off-center shallow donor D$^{-}$ in a spherical quantum dot}


\author{Sergio S. Gomez and Rodolfo H. Romero}

\address{Facultad de Ciencias Exactas, Universidad Nacional del Nordeste, Avenida Libertad 5500 (3400) Corrientes, Argentina}

\begin{abstract}
The binding energy of a negatively charged hydrogenic impurity with on- and off-center position in a spherical Gaussian quantum dot was calculated with the configuraction interaction method. Our calculations show that $E_b$ is always positive for on-center impurities with a maximum near to the radius for one-electron stability of the potential well $R_c$.
For off-center positions the binding energy can assume negative values within a range of the quantum dot radius, thus indicating the instability of the system
\end{abstract}

\end{frontmatter}

% main text
\section{Introduction}
Doping of semiconductor crystallites of nanometer size, or quantum dots (QDs), allows tuning the transport, electric, optical and magnetic properties for the purpose of tailoring proposed quantum devices \cite{Masumoto02}. Incorporation of impurities into QDs provides charge carriers that strongly modifies those properties \cite{Erwin05, Norris08}. 
Neutral and negatively charged shallow donor impurities ($D^0$ and $D^-$ centers) in semiconductors are the analogue of the H atom and the H$^-$ ion in atomic physics, {\em i.e.}, one and two electrons bonded to a positively charged Coulomb center, respectively. In particular, $D^-$ centers are the simplest system where correlation effects can play a role.
The binding energy of a $D^0$ center in QDs has been studied with different confining potential shapes and calculation methods \cite{Zhu90, Kassim07, Silva97, Bose98, Movilla05, Xie08a}. Many of them assume the impurity to be at the center of the QD. Nevertheless, the position of the $D^0$ impurities was shown to strongly affect the binding energy  \cite{Silva97, Bose98, Movilla05, Xie08a}.
Other properties also show such a dependence; for instance, the calculated optical-absorption spectra of homogeneously distributed $D^0$ centers, show an absorption edge associated with transitions involving impurities at the center of the well and a peak related with impurities next to the edge of the dot \cite{Silva97}. 
Also the effect of parabolic confinement on the binding energy of shallow hydrogenic impurities in a spherical QD of a widegap semiconductor, such as GaAs, as a function of the impurity position for different dot sizes, was studied \cite{Bose98}.
The binding energy of an off-center neutral hydrogenic impurity in a spherical quantum dot has been studied by using finite-depth spherical well \cite{Movilla05} and Gaussian confining potentials \cite{Xie08a}. 
The binding energy \cite{Zhu92} and the energy levels of the ground and the excited states of spin-singlet and spin-triplet configurations have been calculated variationally by assuming a square finite-well confining potential \cite{Szafran98}.
%==================================================
%\subsection{On-center $D^-$}
Since the experimental demostration of the existence of built-in $D^-$ centers in doped multiple quantum wells structures \cite{Huant90}, a number of works considered the binding energy of on-center negatively charged impurities, under different confining potentials \cite{Xie99, Pandey04, Gu05, Xie08b, Riva04}.
Xiew proposed a procedure to calculate energy spectrum of $D^-$ centers in disk-like QDs with a parabolic lateral confining potential. He found that there exists a critical radius $R^c$, such that if $R<R^c$ the $D^-$ configuration is stable \cite{Xie99}.
Pandey {\em et al.} studied the dependence of the binding energy of $D^0$ and $D^-$ centres on the confining potencial shape by using the local density approximation \cite{Pandey04}. 
A Gaussian confining potential, having finite depth and range, has been suggested as way to take into account effects of non-parabolicity in the QD potential for both one- and few-electron systems \cite{Adamowski00, Boyacioglu07, Gomez08}. The energy spectra of $D^-$ centres in disk-like Gaussian quantum dots were calculated in Ref. \cite{Gu05}. Recently Xie calculated the binding energy of an on-center $D^-$ donor in a Gaussian potential \cite{Xie08b} and Sahin showed that the use of the exchange and correlation potential is necesary, within the the local density approximation of density functional theory, for obtaining correct results \cite{Sahin08}.
In Ref. \cite{Riva04} the binding energy of an off-center $D^-$ impurity in a two-dimensional parabolic QD was addressed by using finite-difference and fractional dimension methods.\\
To our knowledge, the issue of an off-center $D^-$ donor in a spherical QD has not been addressed. 
Therefore, the purpose of the present work is to study the binding energy of a $D^-$ center in a spherical QD as a function of its position, and to explain this dependence in terms of a simple model.
%###################################################################################
\section{Theory}
We consider two electrons bonded to a shallow donor impurity in a spherical QD of radius $R$ and depth $V_0$. 
The impurity is located at the position ${\bf d}$  and a Gaussian confining potential $V(r) = -V_0 e^{-r^2/2R^2}$ is assumed for the QD.
The Hamiltonian, in the effective mass approximation, can be written as 
%============================================================
\begin{equation}
\label{hamiltonian}
H=\sum_{i=1,2}\left[-\frac{1}{2}\nabla_i^2+V(r_i) +W({\bf d},{\bf r}_i)\right]+ \frac{1}{ r_{12}},
\end{equation}
where $W({\bf d},{\bf r}) = -|{\bf r}-{\bf d}|^{-1}$
is the electron-donor Coulomb potential.
%============================================================
We use the donor Bohr radius $a_D=(\epsilon/m^*)a_{\rm B}$ as the unit of length and the donor effective atomic unit ${\rm a.u.}^*=(m^*/\epsilon^2)$ Hartree as the unit of energy. 
The binding energy of the $D^-$ center is defined as \cite{Zhu92}
\begin{equation}
\label{def Eb}
E_b=E(D^0) + E(e)-E(D^-),
\end{equation}
where $E(D^0)$ is the energy of the neutral impurity $D^{0}$ in the QD, $E(e)$ is the energy of an electron in the QD wihout the impurity, and  $E(D^-)$ is the energy of the $D^-$ in the QD. The energies $E(D^0)$ and $E(e)$ of the one-electron systems are calculated by direct diagonalization. The calculation method was reported elsewhere \cite{Gomez08}.
The energy $E(D^-)$ of the two-electron system were calculated with the configuration interaction (CI) method \cite{Fulde95}, where the eigenvectors of two electron hamiltonian Eq. (\ref{hamiltonian}) are expanded in terms of the two-electron Hartree-Fock ground state and its single and doubly excitated configurations (Slater determinants), expanded in a single-particle Cartesian Gaussian basis set
%\begin{eqnarray}
%\Psi(1,2)=\sum_{i \ell ,j \ell'} c_{i\ell,j\ell'} \varphi^{i}_{\ell}(1)\varphi^{j}_{\ell`}(2)|S,M \rangle 
%\end{eqnarray}
%where $|S,M\rangle$ is an eigenstate of the total spin of the electrons, and the functions $\varphi_\ell^{(i)} $, are cartesian gaussian functions of the form:  
\begin{equation}
\varphi_\ell^{(i)} =  x^m y^n z^p \exp(-\alpha_i r^2),
\end{equation}
where $\ell = m+n+p$ is the angular momentum of the function, and the $\alpha_i$ are properly chosen exponents \cite{Gomez08}. A basis set $4s4p4d$ centered in the QD, and a $8s7p2d$ basis set centered at ${\bf r}={\bf d}$, similar to other previously used for describing the weakly bonded H$^-$ ion and its polarizability, was also added for taking into account the donor center \cite{note_basis}.
The  total spin symmetry of the configurations considered were restricted to $S=0$ as in previous works \cite{Szafran98, Xie99, Gu05}.
A potential depth of $V_0=25$ a.u.$^*$ was kept throughout the present work.
%===================================================================================
\section{Results}
\begin{figure}\caption{\label{E_b}
Binding energy of the Gaussian quantum dot with a negatively charged donor impurity at a distance $d$ from the center for $d=0, 0.3, 0.6$ and $1 a_D$. Energies are given in effective atomic units and distances in donor effective Bohr radius $a_D$. The empty circles represent configuration interaction calculations. The continuous lines are results of the variational model, Eq. (\ref{Eb_m}).}
\begin{tabular}{cc}
\includegraphics[scale=0.45]{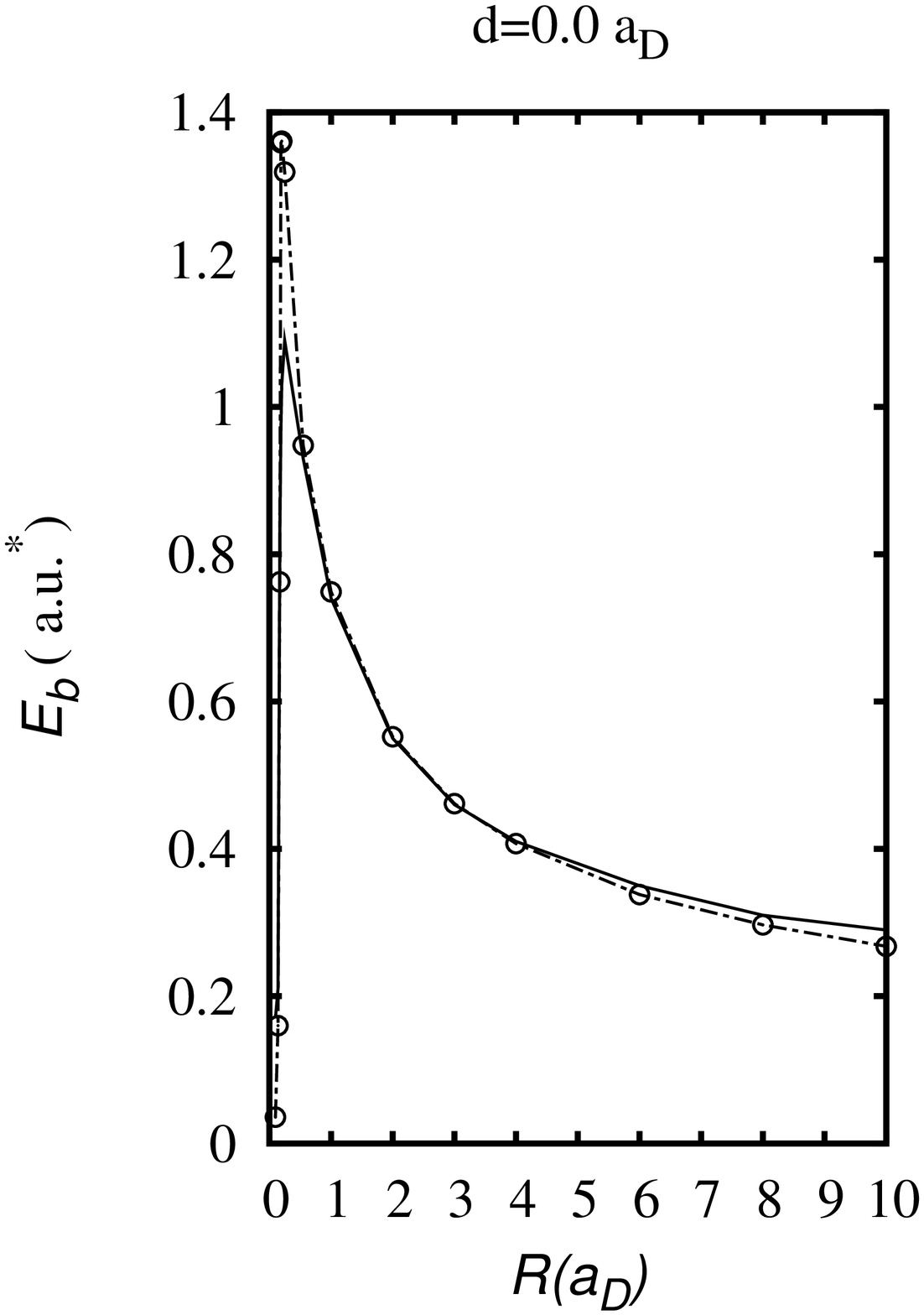} &
\includegraphics[scale=0.45]{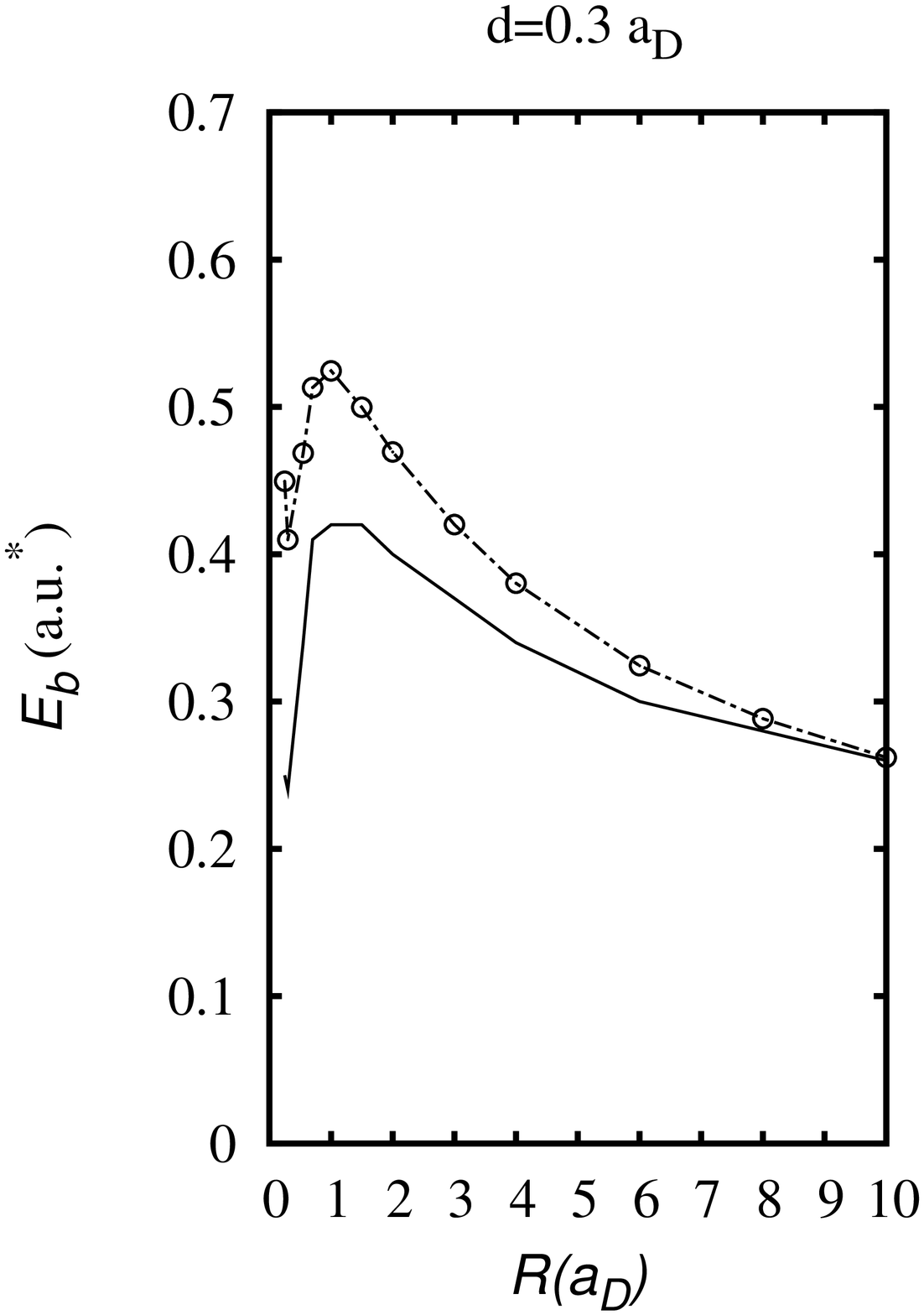} \\
\includegraphics[scale=0.45]{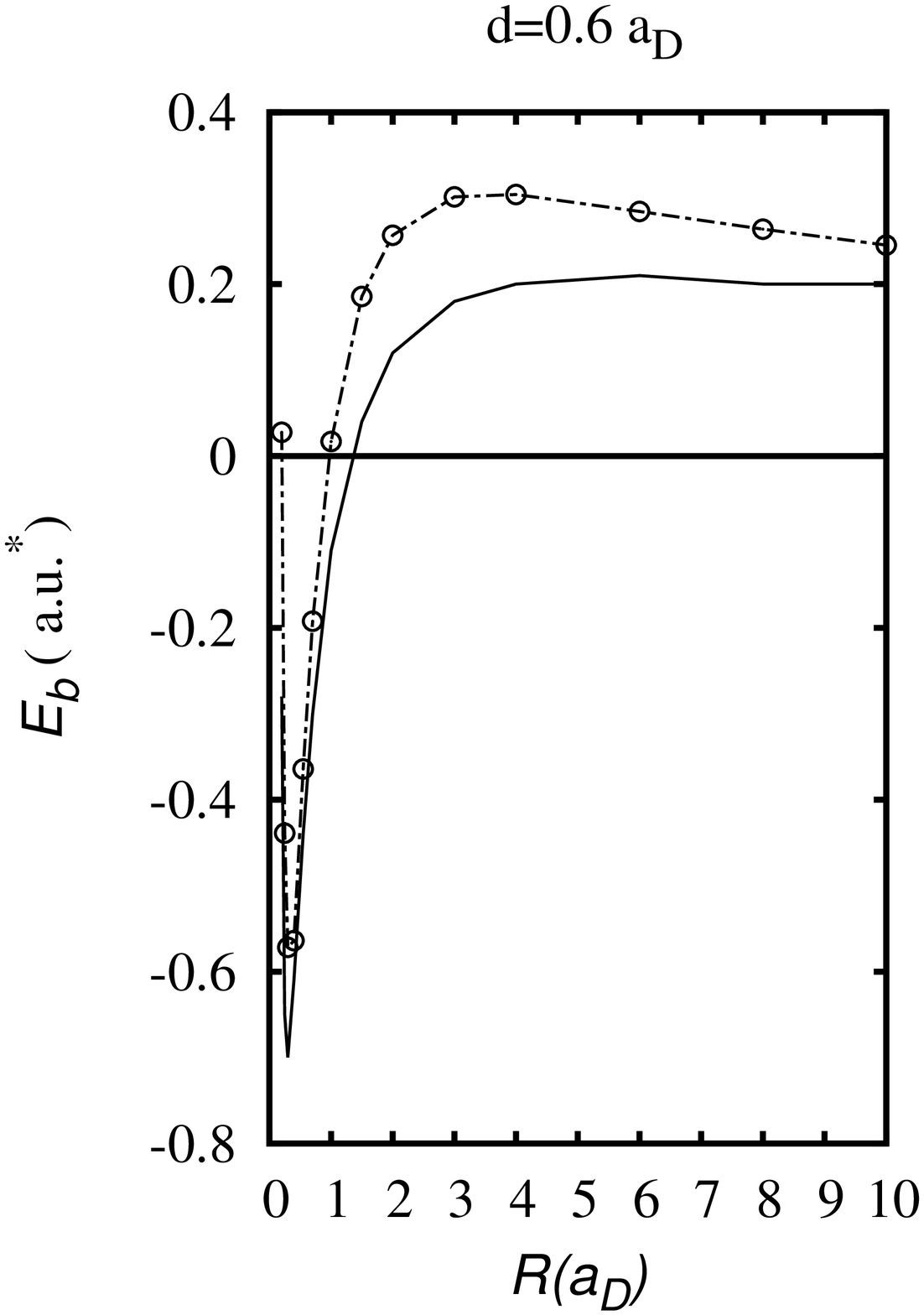} &
\includegraphics[scale=0.45]{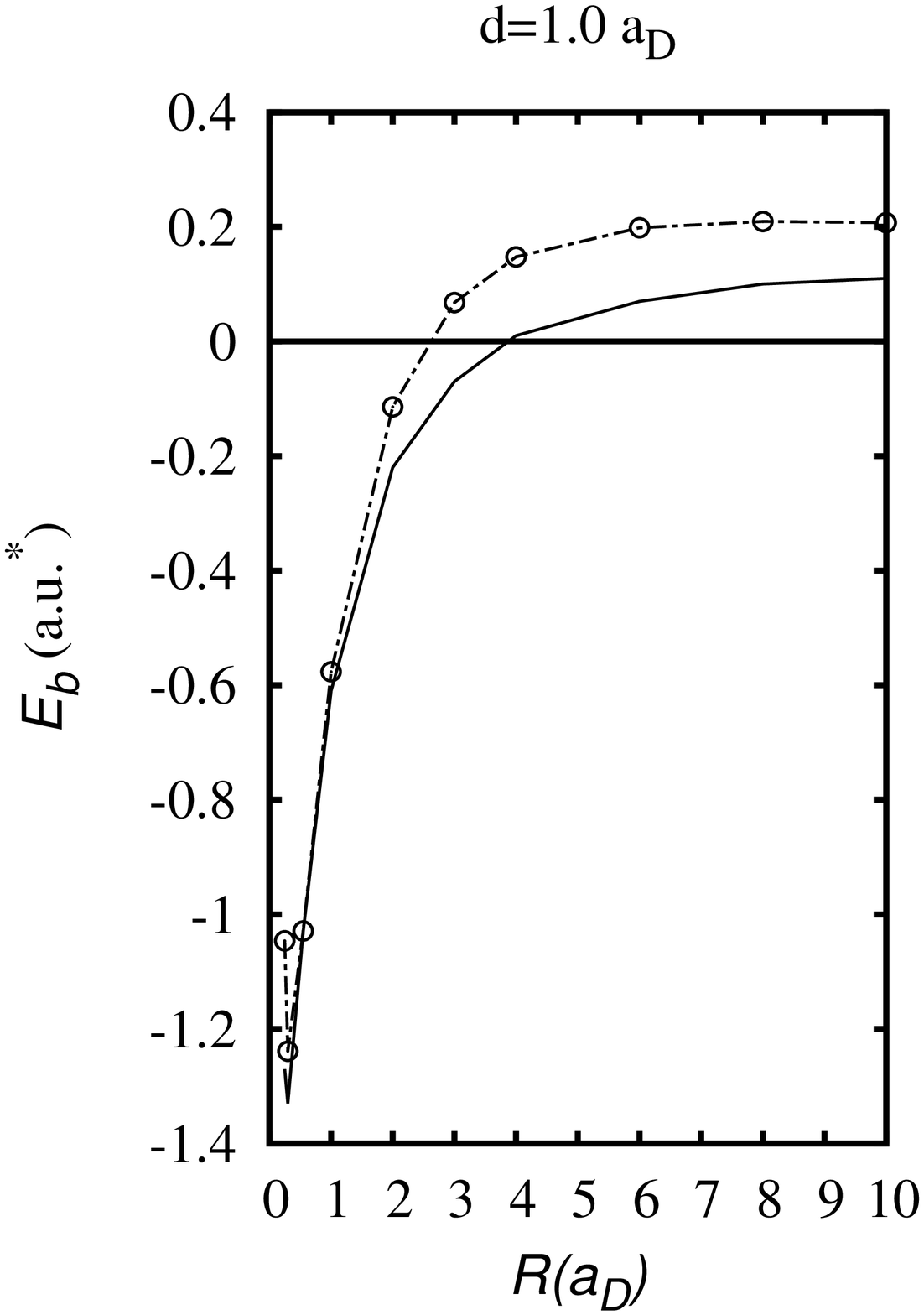} \\
\end{tabular}
%\end{tabular}
\end{figure}

The binding energy, Eq. (\ref{def Eb}), calculated with the CI method is shown in Fig. \ref{E_b} with empty circles as a function of the QD radii $R$. Four impurity positions were considered, namely, $d=0, 0.3, 0.6$ and 1.0 $a_D$. The results show that for on-center position ($d=0$) $E_b$ is always postive with a maximum nearly $R_c\simeq 0.2 a_D$. This maximum binding energy at this critical radius $R_c$, is related to the fact that for every $V_0$, there is a minimum $R_c$ where an electron can be stable in the QD \cite{Gomez08}.
This result is in qualitative agreement with previous works that treated on-center $D^-$ donors \cite{Zhu92, Pandey04, Sahin08}. It should be mentioned, however, that they differ from a recent calculation by Xie \cite{Xie08b}.
At small distances from the potential center ($d=0.3 a_D$), the maximum is less pronounced and there is a minimum at $R_c$. 
For larger values of $d$ (0.6 $a_D$ and $1 a_D$), there still exists a minimum at $R_c$, such that the larger $d$, the more negative the minimum becomes while the maximum becomes flatter.
Also the binding becomes negative for radii $R\sim d$, and positive for $R\gtrsim d$. Hencefore, the larger $d$, the wider the radii range where the binding energy is negative. 
We also preformed Hartree-Fock calculations, not reported here, whose results show a similar trend. The correlation energies were found in the range of $-0.032$ a.u.$^*$ for $R=0.25a_D$, to $-0.037$ a.u.$^*$  for $R=10a_D$, and weakly dependent on the impurity position.

The results can be rationalized as follows. 
For a fixed potential depth and very small radius the effect of the potential becomes negligible because it cannot bind electrons. So, the two electrons are kept bonded due to the impurity Coulomb potential forming a H$^-$ ion.
The same happens for very large radius, where the bottom of Gaussian potential becomes flat and contributtes approximately with a constant potential $-V_0$. Then, $E_b(D^-)\rightarrow E_b({\rm H}^-)=0.0277$ a.u. for both $R\rightarrow 0$ and $R\rightarrow \infty$.

For intermediate radius ($R_c\lesssim R\lesssim d$), where the dot can allocate electrons, the impurity is outside the dot and the system could become instable. 
For very large radius ($R\gg d$), the system behaves like an on-center impurity, thus having a positive binding energy.

\begin{figure}\caption{\label{Eb_mf}
Binding energy of the $D^-$ impurity as a function of the radius of the quantum dot calculated with Eq. (\ref{Eb_m}) for the impurity positions $d=0,\ 0.1,\ 0.2,\ 0.3,\ 0.4,\ 0.6$  and 1 $a_D$. Energies are given in effective atomic units.}
\begin{center}
\includegraphics[scale=1.0]{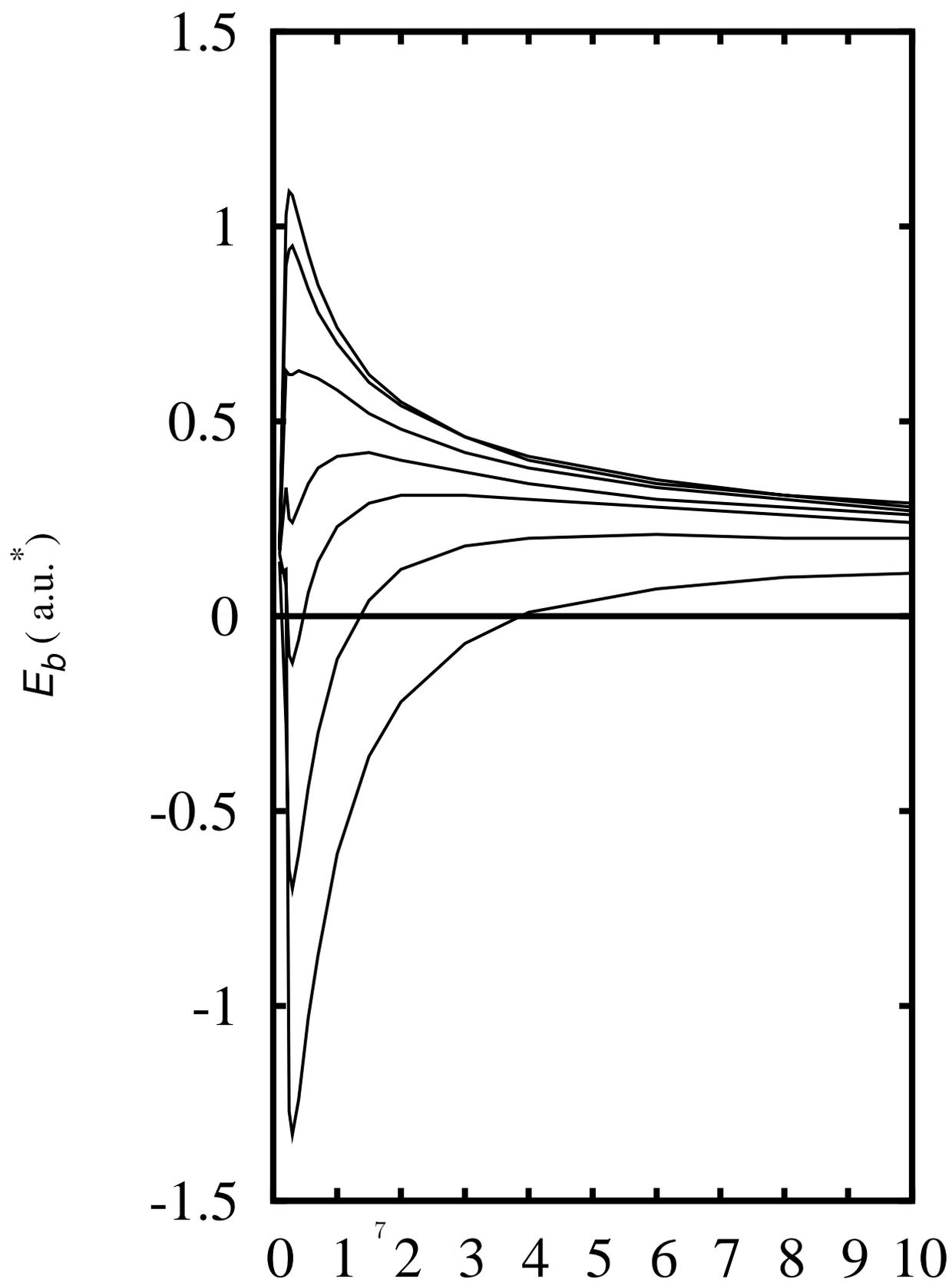}
\end{center}
%\end{tabular}
\end{figure}

%%%%%%%%%%%%%%%%%
A more quantitative explanation of the results can be obtained by using a variational estimate as follows. Consider a normalized $s$-type Gaussian trial function $\varphi_s(r)=(2\alpha/\pi)^{3/4}\exp(-\alpha r^2)$, centered in the QD center.
The energy of the two-electron $D^-$ center can be obtained as the expectation value of the Hartree-Fock Hamiltonian in the spin-singlet trial state $\psi(r_1,r_2)=\varphi_s(r_1)\varphi_s(r_2)$, thus giving
%%%%%%%%%%%%%%%%%
\begin{eqnarray}\label{E_Dm}
E(D^-)= 2\left[\frac{3}{2}\alpha -V_0 \left( \frac{2\alpha}{ 2\alpha + \lambda}\right)^{3/2} -\frac{ {\rm erf}(\sqrt{2\alpha}d)}{d}\right] + 2\sqrt{\frac{\alpha}{\pi}},
\end{eqnarray} 
where the terms within brackets are the expectation value of the kinetic energy $T_\alpha$, confining potential $V_\alpha$ and the impurity potential $W_{\alpha}$ for each electron. The last term is the Coulomb repulsion $J$ between the Gaussian charge densities of each electron.
The optimal exponent $\alpha$ is obtained by minimization of Eq. (\ref{E_Dm}). 
In this way, using the optimal $\alpha$, the ground state energy $E(D^-)$ can be estimated as $E(D^-)= 2(T_\alpha + V_\alpha + W_{\alpha})+J$.
Analogously, $E(D^0)\simeq T_\alpha + V_\alpha + W_{\alpha}$ and $E(e^-)\simeq T_\alpha + V_\alpha$. Hence, the binding energy is approximately given by $E_b(D^-)=-W_{\alpha}-J$, that is,
\begin{eqnarray}\label{Eb_m}
E_b(D^-)=  \frac{ {\rm erf}(\sqrt{2\alpha}d)}{d} - 2\sqrt{\frac{\alpha}{\pi}}.
\end{eqnarray}
Eq. (\ref{Eb_m}) implies that $E_b(D^-)$ depends directly on the interplay between the nuclear attraction and electron-electron interaction. For $d\rightarrow 0$ (the limit of on-center impurity), ${\rm erf}(x)\approx 2x/\sqrt{\pi}$ and Eq. (\ref{Eb_m}) becomes $E_b=2(\sqrt{2}-1)\sqrt{\alpha/\pi}$, thus showing that the on-center binding energy is always positive, in agreement with the CI results presented here. On the other hand, for a fixed $R$, as $d$ increases, ${\rm erf}(\sqrt{2\alpha}d)/d\rightarrow 0$, and $J$ becomes dominant, thus giving $E_b<0$.
The systems for which $E_b<0$ are not stable and is similar to a molecular dissociation process ending up with one electron in a QD and the other in the $D^-$.
The results calculated with Eq. (\ref{Eb_m}) are shown for comparison in Fig. \ref{E_b} with continuous lines. As can be seen, all qualitative features of the CI curves are well reproduced with this simple model.
It is interesting to point out that the use of the Hartree-Fock Hamiltonian gives an electron-electron interaction $2J-K$, where $K$ is the exchange energy such that $K=J$ for the doubly occupied ground state. Thus, Eq. (\ref{E_Dm}) takes into account the exchange energy correctly. Disregarding the exchange energy would imply to add a factor of two to the last term of Eq. (\ref{E_Dm}), and the binding energy for an on-center impurity would give $E_b(D^-)\simeq -W_\alpha-2J\approx 2(\sqrt{2}-2)\sqrt{\alpha/\pi} <0$, in agreement with Ref. \cite{Sahin08}.
Eq. (\ref{Eb_m}) was also used in Fig. \ref{Eb_mf} to show the change in the binding energy as the impurity moves from the center of the QD  up to $d=1a_D$.

%%%%%%%%%%%%%%%%%%%%%%%%%%%%%%%%%%%%%%%%%%%%%%%%%%%%%%%%%%%%%%%%%%%%%%%%%%%%%%%%%%
\section{Concluding remarks}
In summary, we have calculated the binding energy of a negatively charged impurity with on- and off-center position in a spherical gaussian quantum dot with the configuraction interaction method. Our calculations show that $E_b$ is always positive for on-center impurities with a maximum near to the radius for one-electron stability of the potential well $R_c$. As the impurity is displaced off center, the maximum of $E_b$ decreases and a minimum near to $R_c$ appears. For sufficiently large $d$, $E_b$ assumes negative values indicating the instability of the system.  
Our results could be useful for understanding how the binding energy is affected by the breaking of the spherical symmetry of the potential well due to doping in low-dimensional systems.
%###################################################################################

%%%%%%%%%%%%%%%%%%%%%%%%%%%%%%%%%%%%%%%%%%%%%%%%%%%%%%%%%%%%%%%
%%%%%%%%%%%%%%%%%%%%%%%%%%%%%%%%%%%%%%%%%%%%%
\section*{Acknowledgements}
This work has been supported by CONICET (Argentina), Agencia Nacional de Promoci\'on Cient\'{\i}fica y Tecnol\'ogica (ANPCyT, Argentina) and Universidad Nacional del Nordeste under Grants PICTO-204/07 and PI-112/07.
%##################################################################################

\label{}

% \appendix

% \section{}
% \label{}

%#################################################################################

\begin{thebibliography}{00}
\bibitem{note_basis} The basis set used for the hydrogenic impurity was constructed by geometric continuation of the Sadlej basis set for hydrogen adding $2s$ (exponents 0.0108 and 0.0036) and $3p$ diffuse functions (exponents 0.0108, 0.007 and 0.0035) and $2d$ (exponents 0.5166, 0.1722) as in Ref. \cite{Fowler84}.
\end{thebibliography}
\end{document}